# A New Procedure for the Photometric Parallax Estimation

**S. KARAALİ[1], Y. KARATAŞ[1], S. BİLİR[1], S. G. AK[1], and E. HAMZAOĞLU[2]**

[1]*Istanbul University, Science Faculty, Department of Astronomy and Space Sciences, 34452 Istanbul – TURKEY*

[2] *Istanbul Commerce University, Faculty of Engineering and Design, Ragıp Gümüşpala Cad. No: 84, 34378 Istanbul - TURKEY*

E-mail: karsa@istanbul.edu.tr

**Abstract**

We present a new procedure for photometric parallax estimation. The data for 1236 stars provide calibrations between the absolute magnitude offset from the Hyades main-sequence and the UV-excess for different $(B-V)_o$ colour-index intervals, i.e.: (0.3 0.4), (0.4 0.5), (0.5 0.6), (0.6 0.7), (0.7 0.8), (0.8 0.9), (0.9 1.0), and (1.0 1.1). The mean difference between the original and estimated absolute magnitudes and the corresponding standard deviation are rather small, +0.0002 and ±0.0613 mag. The procedure has been adapted to the Sloan photometry by means of colour equations and applied to a set of artificial stars with different metallicity. The comparison of the absolute magnitudes estimated by the new procedure and the canonical one indicates that a single colour-magnitude diagram does not supply reliable absolute magnitudes for stars with large range of metallicity.

**1. Introduction**

Stellar kinematics and metallicity are two primary means to deduce the history of our Galaxy. However, such topics can not be achieved without stellar distances. The distance to a star can be evaluated by trigonometric or photometric parallaxes. Trigonometric parallaxes are only available for nearby stars where Hipparcos (ESA, 1997) is the main supplier for the data. For stars at large distances, the use of photometric parallaxes is unavoidable. In other words the study of the Galactic structure is strictly tied to precise determination of absolute magnitudes.

Different methods could be found for absolute magnitude determination. The method used in the Strömgren's uvby-β (Nissen & Schuster 1991) and in the UBV (Laird et al. 1988, hereafter LCL) photometry depends on the absolute magnitude offset from a standard main-sequence. In recent years the derivation of absolute magnitudes has been carried out by means of colour-absolute magnitude diagrams of some specific clusters whose metal abundances are generally adopted as the mean metal abundance of a Galactic population, such as thin, thick disks, and halo. The studies of Phleps et al. (2000) and Chen et al. (2001) can be given as examples. A slight different approach is that of Siegel et al. (2002) where two relations, one for stars with solar-like abundances and another one for the metal-poor stars were derived between $M_R$ and the colour-index R-I, where $M_R$ is the absolute magnitude in the R filter of Johnson system. For a star of given metallicity and colour-absolute magnitude can be estimated by linear interpolation of two ridgelines and by means of linear extrapolation beyond the metal-poor ridgeline.

We strongly believe to contribute to this important topic by modifying the method of LCL and by adapting it to the Sloan photometry. LCL used the equation :

$$M_V(Hyades) = 5.64(B-V)_o + 1.11 \qquad (1)$$

for the fiducial main-sequence of Hyades as a standard main-sequence and derived the metallicity-dependent offset from this equation :

$$\Delta M_V^H = \frac{2.31 - 1.04(B-V)_o}{1.594}(-0.6888\delta + 53.14\delta^2 - 97.004\delta^3) \qquad (2)$$

LCL state the calibration to be valid for $\delta \leq 0.25$, which is equal to [Fe/H] = -1.75 dex, according to the Carney (1979) transformation of $\delta$ into [Fe/H] :

$$[Fe/H] = 0.11 - 2.90\delta - 18.68\delta^2 \qquad (3)$$

Moreover, LCL give an equation for direct absolute magnitude derivation for extreme metal-poor stars :

$$M_V(B-V) = 4.60(B-V)_o + 3.36 + 1.67(\delta - 0.25) \qquad (4)$$

As can be revealed from these explanations, the method of LCL is based on the fact that absolute magnitude (and metallicity) is a function of UV-excess, additional to colour-index. UV-excess is usually defined as the de-reddened (U-B) colour-index difference between a star and a Hyades star of equal (B-V)$_o$. The U band is centered at a wavelength where metallicity effect is efficient, hence a star with bright U magnitude, i.e.: a relatively metal-poor star, is absolutely faint relative to a Hyades star of equal (B-V)$_o$.

We considered the possibility to calibrate the absolute magnitude offset from the update Hyades sequence (see Appendix),

$$M_V(Hyades) = -1.48739(B-V)_o^2 + 7.70982(B-V)_o + 0.331195 \qquad (1')$$

using only δ for different (B-V)$_o$ colour-indices without any restriction for metallicity. This is the main scope of this work. We will show in the following sections that such an approach provides more precise absolute magnitudes than those of LCL. In section 2 we present the data used for calibration and in section 3 the procedure used for calibration is given. The extension of this procedure to the Sloan photometry is given in section 4 and in section 5 a detailed discussion is provided.

## 2. The Data

The V, B-V, U-B, and E(B-V) photometric data used in this paper and the star distance *d* are taken from Ryan (1989). For any star the following reductions have been applied :

$$\begin{aligned}
(B-V)_o &= (B-V) - E(B-V) \\
(U-B)_o &= (U-B) - 0.72E(B-V) + 0.05E^2(B-V) \\
M_V &= V - 3.1E(B-V) + 5 - 5\log d \\
\delta(U-B) &= (U-B)_H - (U-B)_o
\end{aligned} \qquad (5)$$

(U-B)$_H$ is the de-reddened (U-B) colour-index of a dwarf star of the Hyades cluster with the same (B-V)$_o$ of the star considered. We indicate with $\Delta M_V^H$ the absolute magnitude difference between a star and a Hyades star of equal (B-V)$_o$ and with $\delta_{0.6}$ the normalized UV-excess of the star considered (see Table 1). i.e.: $\delta_{0.6}$ is the de-reddend (U-B) colour-index difference between two stars just quoted and necessary coefficient used here is given by Sandage (1969). This procedure is necessary for the equivalence of UV-excess of two stars of the same metal-abundance, one with any (B-V)$_o$ and another one with (B-V)$_o$ = 0.6 where the latter is adopted as a reference colour-index for this reduction (Sandage, 1969). Contrary to Laird et al. (1988) who gave relations as a function of both colour-index (B-V)$_o$ and $\delta_{0.6}$ (equation 2) we prefered to plot $\Delta M_V^H$ as a function of only $\delta_{0.6}$ for different (B-V)$_o$ intervals i.e.: 0.3 < (B-V)$_o$ ≤ 0.4; 0.4 < (B-V)$_o$ ≤ 0.5; 0.5 < (B-V)$_o$ ≤ 0.6; 0.6< (B-V)$_o$ ≤ 0.7; 0.7 < (B-V)$_o$ ≤ 0.8; 0.8< (B-V)$_o$ ≤ 0.9; 0.9 < (B-V)$_o$ ≤ 1.0; 1.0 < (B-V)$_o$ ≤ 1.1. This approach improves significantly the calibrations with respect to those of LCL, as explained in the following sections.

### 3. Photometric Parallaxes

#### 3.1. Calibration of absolute magnitude as a function of UV-excess

We used $\Delta M_V^H = M_V(*) - M_V(H)$ and the $\delta_{0.6}$ listed in Table 1 for a third-degree polynomial fitting for each (B-V)$_o$ interval cited in section 2, where $M_V(H)$ and $M_V(*)$ are the absolute magnitudes of a Hyades star (evaluated by equation 1') and of a programme star of equal (B-V)$_o$, respectively. Stars are separated into different bins in $\delta_{0.6}$ with range $\Delta\delta_{0.6} = 0.05$ mag. in order to take into account all the programme stars and to provide reliable statistics. The number of bins is 6 for the bluest and reddest intervals of 0.3 <(B-V)$_o \leq 0.4$ and $1.0 <$ (B-V)$_o \leq 1.1$ and lie between 9 and 12 for the other six colour-index intervals. The mean of $\delta_{0.6}$ and $\Delta M_V^H$ are evaluated for each bin except one bin in each interval of $0.7 <$ (B-V)$_o \leq 0.8$; $0.8<$ (B-V)$_o \leq 0.9$; $0.9 <$ (B-V)$_o \leq 1.0$ and $1.0 <$ (B-V)$_o \leq 1.1$, which have relatively extreme $\delta_{0.6}$ or $\Delta M_V^H$ values. According to this criterion eight stars were excluded from the analysis (see Table 2). Then $\Delta M_V^H$ was plotted versus $\delta_{0.6}$ (Fig.1) and a third-degree polynomial was fitted for each set of data,

$$\Delta M_V^H = a_3 \delta_{0.6}^3 + a_2 \delta_{0.6}^2 + a_1 \delta_{0.6} + a_o \qquad (6)$$

The coefficients $a_i$ of this equation are given in Table 3 as a function of (B-V)$_o$. One notices two important points in Fig.1. First, a large scattering between the curves exist and, second, contrary to expectations neither of the curves converge towards the origin. This means that a star with $\delta_{0.6} = 0$, corresponds to an absolute magnitude of a Hyades star, would have a value for $\Delta M_V^H$ different than zero hence, different absolute magnitude than the Hyades star, according to the definition of $\Delta M_V^H$. This is a contradiction. The curves should pass through the origin to avoid this discrepancy. Table 3 shows that all the zero points are larger than $\Delta M_V^H = 0.2$. It increases from $\Delta M_V^H = 0.24$ for the interval $0.3 <$ (B-V)$_o \leq 0.4$ to a maximum value of $\Delta M_V^H = 0.35$ at the interval $0.6 < (B-V)_o \leq 0.7$ and declines to lower values such as $\Delta M_V^H = 0.29$ for the interval $1.0 < (B-V)_o \leq 1.1$.

We would like to quote the work of Cameron (1985) who discussed the same relation, i.e.: $\Delta M(V)$ versus $\delta_{0.6}$. We fitted a third degree polynomial for his data (Table 2 and Fig.6) with a constant term of -0.1663 which is absolutely equal and almost half of the mean of the constant terms in our work. The work of Cameron (1985) also indicates that $\delta_{0.6} = 0$ does not imply $\Delta M(V) = 0.0$.

### 3.2. Normalization of the Hyades main sequence

The discrepancy mentioned above can be minimized by normalization of the Hyades main-sequence. In other words the absolute magnitude corresponding to the Hyades fiducial main-sequence needs an increment to limit the constant term in equation (6). Table 4 gives the Hyades absolute magnitudes evaluated by equation (1') $M_V^H(ev)$ and the adopted one, i.e.: $M_V^H(ad) = M_V^H(ev) + a_o$ where $a_o$ is the corresponding zero point in equation (6). The adopted absolute magnitudes are plotted against the mean $(B-V)_o$ for each interval and the following quadratic equation has been fitted (Fig.2).

$$M_V^H(nor) = -2.1328(B-V)_o^2 + 8.6803(B-V)_o + 0.305 \qquad (7)$$

This is the normalized colour-magnitude equation for the Hyades main-sequence used in the derivation of photometric parallaxes.

### 3.3. Final equations for photometric parallaxes

After normalization of the absolute magnitudes for the Hyades main-sequence, the difference in absolute magnitude between a star and a Hyades star of equal $(B-V)_o$ i.e.: $\Delta M_V^H(nor)$ is re-evaluated and used in final equations for photometric parallaxes (see Table 5). The procedure is the same as in section 3.1. The mean of $\Delta M_V^H(nor)$ for each bin is given in the fourth column of Table 2. The plot of the $\Delta M_V^H(nor)$ against $\delta_{0.6}$ for each $(B-V)_o$ interval yields the following third-degree polynomial :

$$\Delta M_V^H(nor) = b_3\delta_{0.6}^3 + b_2\delta_{0.6}^2 + b_1\delta_{0.6} + b_o \qquad (8)$$

The coefficients $b_i$ of this equation are given in Table *6* and the plots are shown in Fig.3. The curves in Fig.3 exhibit a different appearance than the corresponding ones in Fig.1. Dispersion of the curves in Fig.3 is lesser and, now all the curves pass almost through the origin (see the term $b_o$ in Table *6*). We used equation (8) to evaluate the absolute magnitude offset $\Delta M_V^H(nor)$ for all the programme stars and to estimate their absolute magnitudes $M_V(est)$ by the definition of the offset i.e.: $\Delta M_V^H(nor) = M_V(est) - M_V^H(nor)$. Surprisingly, the differences between the estimated and original absolute magnitudes are rather small. The mean of these differences for each $(B-V)_o$ - interval is almost zero and their standard deviations are only few percent. However, this is not the result for the procedure applied by LCL (Table 7 and Fig.4, see section 5 for detailed discussion).

The evolutionary effect has not been considered above. However, the $(U-B)_o$ versus $(B-V)_o$ sequence slightly varies as a function of the gravity. Therefore for stars close to the end of the main sequence TAMS, the estimate of real $\delta_{0.6}$ is smaller. We used the Yale isochrones of Yi et al. (2001) for the following chemical composition and checked the size of the errors introduced by evolutionary effects between the ZAMS and TAMS (10 Gyr): Y = 0.27 and Z = 0.02 ([Fe/H] = +0.05 dex). For this sample, $<\delta_{0.6}>$= 0.0 and 0.02 for $0.81 < (B-V)_o \leq 1.00$ and $0.7 < (B-V)_o \leq 0.81$ respectively. The effect of this difference in $\Delta M_V^H(nor)$ for a star with $\delta_{0.6} = 0.2$ ([Fe/H] = -1.2 dex) is 8%.

### 4. Extension of the procedure to the Sloan photometry

### 4.1. Transformation of the normalized UV-excess from UBV to the Sloan photometry and new metallicity calibration

The following colour equations of Fukugita et al. (1996) provide a relation between the normalized UV-excesses for UBV and Sloan photometries and new metallicity calibration for the latter photometric system :

$$(g'-r')_o = 1.05(B-V)_o - 0.23$$
$$(u'-g')_o = 1.38(U-B)_o + 1.14 \qquad (9)$$

Let us write the second of equations (9) for two stars with the same (B-V)$_o$ (or equivalently (g'-r')$_o$), i.e.: for a Hyades star (H) and for a star (*) whose UV-excess is normalized to :

$$(u'-g')_H = 1.38(U-B)_H + 1.14$$
$$(u'-g')_* = 1.38(U-B)_* + 1.14 \qquad (10)$$

Then, the UV-excess for the star in question relative to the Hyades star is :

$$(u'-g')_H - (u'-g')_* = 1.38\left[(U-B)_H - (U-B)_*\right] \qquad (11)$$

or in standard notation :

$$\delta(u'-g') = 1.38\delta(U-B) \qquad (12)$$

If we apply this equation to a star with colour-index (B-V)$_o$ = 0.6, corresponding to (g'-r')$_o$ = 0.4, we obtain :

$$\delta(u'-g')_{0.4} = 1.38\delta(U-B)_{0.6} \qquad (13)$$

for the relation between the normalized UV-excesses in the UBV and the Sloan systems. From this equation we obtain :

$$\delta(U-B)_{0.6} = 0.725\delta(u'-g')_{0.4} \qquad (14)$$

which yields a new metallicity calibration for the Sloan photometry by its substitution in the following equation which covers a large range of metallicity, i.e.: -2.75 ≤ [Fe/H] ≤ 0.2 dex (Karaali et al., 2003) :

$$[Fe/H] = 0.10 - 2.76\delta_{0.6} - 24.04\delta_{0.6}^2 + 30.00\delta_{0.6}^3 \qquad (15)$$

Hence, the new metallicity calibration for the Sloan photometry is obtained as follows:

$$[Fe/H] = 0.10 - 2.00\delta_{0.4} - 12.64\delta_{0.4}^2 + 11.43\delta_{0.4}^3 \qquad (16)$$

Finally, we can show that the coefficients given by Sandage (1969) for the UBV photometry can also be used for the normalization of the UV-excesses in the Sloan photometry. Let's take another star with any B-V (or equivalent g'-r') but with the same metallicity as the first star. The relation between its normalized UV-excesses in the two systems would be as equation (12). Hence, from (12) and (13) we obtain:

$$\delta(u'-g')_{0.4} / \delta(u'-g') = \delta(U-B)_{0.6} / \delta(U-B) = f \qquad (17)$$

where $f$ is the UV-excess normalized factor.

### 4.2. Photometric parallaxes for the Sloan photometry

The procedure in section (3.3) can be adopted for photometric parallax derivation also for the Sloan photometry by using the colour equations and the relation between the normalized UV-excesses in two systems as mentioned above. First, we draw $(B-V)_o$ from the first equation in (9) i.e.:

$$(B-V)_o = 0.952(g'-r')_o + 0.219 \qquad (18)$$

and then substitute it into (7) for normalization of the Hyades main-sequence in the Sloan photometry as follows:

$$M_{g'}^H(nor) = -1.9330(g'-r')_o^2 + 7.3742(g'-r')_o + 2.1036 \qquad (19)$$

Bearing in mind that the offsets from the fiducial sequence of Hyades in two systems are equal, the one for the Sloan photometry can be derived by replacing the equivalence of $\delta(U-B)_{0.6}$ in (14) into (8). The following result is obtained:

$$\Delta M_{g'}^H(nor) = c_3\delta_{0.4}^3 + c_2\delta_{0.4}^2 + c_1\delta_{0.4} + c_o \qquad (20)$$

### 4.3. Comparison the absolute magnitudes derived by the new procedure and the colour-magnitude diagram of a specific cluster

We compared the absolute magnitudes derived by the new procedure and the colour- magnitude diagram of cluster M13 used for the photometric parallax estimation for halo dwarfs (cf. Chen et al. 2001) just for an example. One can extend this comparison to the other components of the Galaxy. The work is carried out as follows. First, we used the UBV data of Richer & Fahlman (1986) and evaluated the $(g'-r')_o$ and the $M(g')$ absolute magnitudes for the main-sequence of M13 via the first of equations (9) and the following colour-equation which is adopted from Fukugita et al. (1996) respectively :

$$M(g') = M(V) + 0.56\,(B-V)_o - 0.12 \qquad (21)$$

The $(g'-r')_o$ and $M(g')$ data thus obtained (Table 9) transform the main-sequence of cluster M13 from UBV to the Sloan photometry :

$$M_{g'}(M13) = 11.442\,(g'-r')_o^3 - 25.292\,(g'-r')_o^2 + 21.599(g'-r')_o + 0.8621 \qquad (22)$$

Equation (22) yields direct absolute magnitude estimation for metal-poor stars such as halo dwarfs.

As a second step, we adopted seven sample of artificial stars with $(g'-r')_o$ between 0.20 and 0.50 but with different metallicities, and evaluated absolute magnitudes for them by using equation (19) and the related one in (20). The selection of this colour-index interval is due to the work of Chen et al. (2001). These authors assumed that stars fainter than g' ~ 18 mag. with $0.20 \leq (g'-r')_o \leq 0.50$ belong to the halo population and used the colour-magnitude diagram of cluster M13, without any metallicity restriction, for their absolute magnitude determination. Whereas we adopted different metallicities for different sample to reveal the difference between two procedures. As it is easier to derive the metallicity by the normalized UV-excess, we adopted $\delta_{0.4}$ = 0.00; 0.10; 0.20; 0.30; 0.40; 0.50 and 0.60 respectively which correspond to the metallicities [Fe/H] = 0.10; -0.21; -0.71; -1,33; -1.99; -2.63 and -3.18 dex. Table 10 gives the full set of $(g'-r')_o$ cited, the corresponding normalized absolute magnitudes for the Hyades main-sequence ($M_{g'}^{H}(nor)$), the offsets from this sequence ($\Delta M_{g'}^{H}(nor)$) and the absolute magnitudes ($M(g')$) in question.

Finally, we evaluated another set of absolute magnitudes by means of equation (22) and compared them with the absolute magnitudes in seven sets mentioned above (Table 11). The mean of the differences between the absolute magnitudes derived by the new procedure and those evaluated by means of equation (22) i.e.: $<\Delta M> \equiv M_{g'}(M13) - M_{g'}(\delta_{0.4})>$ are larger for relatively metal-rich stars as expected and least for [Fe/H] ≈ -2 dex. The following third-degree polynomial fits well the couple <ΔM> and $\delta_{0.4}$ (Fig.5) :

$$\delta_{0.4} = -0.2305 <\Delta M>^3 + 0.5374 <\Delta M>^2 - 0.6575 <\Delta M> + 0.4369 \quad (23)$$

It also reveals that <ΔM> = 0 for $\delta_{0.4}$ = 0.4369 or [Fe/H] = -2.23 dex. The result indicates that a colour-magnitude diagram with metallicity less than the one for M13 ([Fe/H] ≈ -1.4 dex) is more appropriate for the photometric parallax estimation for metal-poor stars, in deep surveys such as SDSS (as explained in the following section).

## 5. Discussion

We used the high-precision UBV data of Ryan (1989) for absolute magnitude estimation. Although LCL had already derived two equations, one for stars with metallicity [Fe/H] ≥ -1.75 dex and another for extreme metal poor stars (equations 2 and 4 respectively), both equations are functions of (B-V)$_o$ and of the normalized $\delta_{0.6}$ UV-excess. However, as it can be derived from Fig.1 and Fig.3, the offset from the fiducial main-sequence of Hyades behaves differently for different colour-index intervals, confirming the necessity of different equations for different (B-V)$_o$ intervals.

As claimed by LCL, they forced their calibration in order to pass through the zero point, thus supplying the Hyades absolute magnitudes for $\delta_{0.6}$ = 0. In this study we used the update data (see Appendix) and obtained a quadratic equation for the Hyades sequence.

Though, our calibration does not pass through the zero point either. Hence, we normalized the fiducial main-sequence of Hyades. This approach supplies absolute magnitudes almost equal to the Hyades absolute magnitudes for $\delta_{0.6} = 0$, for all $(B-V)_o$ intervals.

The comparison of the estimated absolute magnitudes with the original ones confirm the accuracy of our calibration. The mean of the differences of absolute magnitudes for each $(B-V)_o$ interval is almost zero and their standard deviations are only few percent (Table 7). The mean difference for stars with $0.3 < (B-V)_o \leq 1.1$ and the corresponding standard deviation are $+0.0002$ and $\pm 0.0613$ mag., respectively. Moreover, the plot of these differences versus the original absolute magnitudes shows that most of the stars lie within the interval $-0.1 < \Delta M(V) < +0.1$ (Fig.4a). Whereas the comparison of the absolute magnitudes estimated by LCL with the original ones gives larger means and standard deviations (Table 7). Mean difference and the corresponding standard deviation for all stars are $-0.0151$ and $\pm 0.4782$ mag., respectively, rather different values than those from the new procedure. Finally, Fig.4b also demonstrates the large range of the absolute magnitude differences for LCL, i.e.: the majority of stars lie within $-0.5 < \Delta M(V) < +0.5$ and there are about one hundred stars with larger differences additionally.

The colour-equations of Fukugita et al. (1996) provide a new metallicity calibration for the Sloan photometry. This has been carried out by the relation of normalized UV-excesses in the UBV and Sloan photometric systems, i.e.: by substituting $\delta(U-B)_{0.6} = 0.725\ \delta(g'-r')_{0.4}$ into the metallicity calibration of Karaali et al. (2003). Same substitution into the equations (8) transforms the offset from the fiducial main-sequence of Hyades from UBV to Sloan photometry (equations 20) and finally the combination of (19) and (20) provides absolute magnitude estimation for the Sloan photometry.

We applied the new procedure to a set of artificial stars with $(g'-r')_o$ between 0.20 and 0.50, and compared the absolute magnitudes derived for seven different metallicities with the absolute magnitudes evaluated by means of the colour-magnitude diagram of M13. This is an example to see how coincident the present approach and the canonical one are. The mean of the differences between the absolute magnitudes derived by the new procedure and the canonical way is large for relatively metal-rich stars, is zero for the metallicity [Fe/H] = -2.23 dex and has a large range extending from $\langle \Delta M \rangle = 1.269$ to $\langle \Delta M \rangle = -0.186$. It is surprising that the coincidence occurs for the metallicity of M92 but not for the metallicity of M13 ([Fe/H] = -1.4 dex). One can argue that the metal-rich stars are not efficient in the deep surveys. However, the range of <ΔM> extends from +0.4 to –0.2 even for the metallicity range from –1.0 to –3.0 dex, which is dominated by Pop II stars. Additionally, the standard deviations (Table 11) for the seven comparisons mentioned above are larger than σ = ±0.2 mag., resulting in an extra internal error in absolute magnitude estimation. The combination of these effects encourages us to claim that a single colour-magnitude diagram does not supply reliable absolute magnitudes for stars with a large range of metallicity. On the other hand, the small scattering of the differences between the original and the estimated absolute magnitudes for the UBV photometry confirms the significant improvement of the new procedure with respect to the LCL. One finally, regarding to the colour-equations of Fukugita et al. (1996) we argue that the new procedure can also be applied extensively and efficiently to SDSS (and to other systems, using appropriate colour-equations).

**Table 2.** Normalized UV-excesses ($\delta_{0.6}$) and two sets of absolute magnitude differences, i.e.: $<\Delta M_1> \equiv \Delta M_V^H$ and $<\Delta M_2> \equiv \Delta M_V^H$ (nor) defined in sections 3.1 and 3.2 respectively, in different bins for stars in eight (B-V)$_o$ colour index intervals (a) (0.3 0.4), (b) (0.4 0.5), (c) (0.5 0.6), (d) (0.6 0.7), (e) (0.7 0.8), (f) (0.8 0.9), (g) (0.9 1.0), and (h) (1.0 1.1). N is the total number of stars in each bin.

| (a) 0.3 < (B-V)$_o$ ≤ 0.4 | | | | | (b) 0.4 < (B-V)$_o$ ≤ 0.5 | | | | |
|---|---|---|---|---|---|---|---|---|---|
| $\delta_{0.6}$-interval | $<\delta_{0.6}>$ | $<\Delta M_1>$ | $<\Delta M_2>$ | N | $\delta_{0.6}$-interval | $<\delta_{0.6}>$ | $<\Delta M_1>$ | $<\Delta M_2>$ | N |
| (-0.025 +0.025] | -0.010 | 0.213 | -0.046 | 1 | (-0.125 -0.075] | -0.110 | -0.121 | -0.393 | 1 |
| (+0.025 +0.075] | 0.045 | 0.364 | 0.115 | 2 | (-0.075 -0.025] | -0.030 | 0.097 | -0.197 | 1 |
| (+0.175 +0.225] | 0.200 | 0.907 | 0.660 | 2 | (-0.025 +0.025] | 0.003 | 0.290 | 0.006 | 6 |
| (+0.225 +0.275] | 0.260 | 1.249 | 0.992 | 4 | (+0.025 +0.075] | 0.055 | 0.552 | 0.269 | 6 |
| (+0.275 +0.325] | 0.304 | 1.332 | 1.081 | 25 | (+0.075 +0.125] | 0.096 | 0.750 | 0.469 | 7 |
| (+0.325 +0.375] | 0.340 | 1.384 | 1.137 | 11 | (+0.125 +0.175] | 0.159 | 1.102 | 0.814 | 15 |
| | | | | | (+0.175 +0.225] | 0.204 | 1.230 | 0.947 | 59 |
| | | | | | (+0.225 +0.275] | 0.250 | 1.360 | 1.078 | 97 |
| | | | | | (+0.275 +0.325] | 0.301 | 1.525 | 1.224 | 43 |
| | | | | | (+0.325 +0.375] | 0.339 | 1.609 | 1.330 | 9 |
| | | | | | (+0.375 +0.425] | 0.390 | 1.698 | 1.435 | 1 |
| (c) 0.5 < (B-V)$_o$ ≤ 0.6 | | | | | (d) 0.6 < (B-V)$_o$ ≤ 0.7 | | | | |
| $\delta_{0.6}$-interval | $<\delta_{0.6}>$ | $<\Delta M_1>$ | $<\Delta M_2>$ | N | $\delta_{0.6}$-interval | $<\delta_{0.6}>$ | $<\Delta M_1>$ | $<\Delta M_2>$ | N |
| (-0.075 -0.025] | -0.060 | -0.016 | -0.327 | 2 | (-0.175 -0.075] | -0.153 | -0.393 | -0.726 | 3 |
| (-0.025 +0.025] | 0.015 | 0.394 | 0.077 | 2 | (-0.075 -0.025] | -0.028 | 0.215 | -0.119 | 6 |
| (+0.025 +0.075] | 0.053 | 0.633 | 0.317 | 11 | (-0.025 +0.025] | 0.013 | 0.439 | 0.104 | 11 |
| (+0.075 +0.125] | 0.104 | 0.826 | 0.507 | 35 | (+0.025 +0.075] | 0.052 | 0.593 | 0.261 | 26 |
| (+0.125 +0.175] | 0.153 | 1.057 | 0.743 | 55 | (+0.075 +0.125] | 0.102 | 0.834 | 0.502 | 51 |
| (+0.175 +0.225] | 0.202 | 1.240 | 0.928 | 67 | (+0.125 +0.175] | 0.150 | 1.052 | 0.720 | 61 |
| (+0.225 +0.275] | 0.250 | 1.440 | 1.129 | 71 | (+0.175 +0.225] | 0.202 | 1.240 | 0.907 | 39 |
| (+0.275 +0.325] | 0.292 | 1.588 | 1.275 | 24 | (+0.225 +0.275] | 0.250 | 1.416 | 1.085 | 26 |
| (+0.325 +0.375] | 0.337 | 1.740 | 1.429 | 3 | (+0.275 +0.325] | 0.297 | 1.560 | 1.230 | 10 |
| | | | | | (+0.325 +0.425] | 0.353 | 1.571 | 1.239 | 3 |

**(Table 2 cont.)**

| (e) $0.7 < (B-V)_o \leq 0.8$ | | | | | (f) $0.8 < (B-V)_o \leq 0.9$ | | | | |
|---|---|---|---|---|---|---|---|---|---|
| $\delta_{0.6}$-interval | $<\delta_{0.6}>$ | $<\Delta M_1>$ | $<\Delta M_2>$ | N | $\delta_{0.6}$-interval | $<\delta_{0.6}>$ | $<\Delta M_1>$ | $<\Delta M_2>$ | N |
| (-0.200  -0.125] | ----- | ----- | ----- | 2 | (-0.175  -0.125] | -0.154 | -0.228 | -0.561 | 5 |
| (-0.125  -0.075] | -0.089 | -0.104 | -0.442 | 8 | (-0.125  -0.075] | -0.093 | 0.020 | -0.309 | 14 |
| (-0.075  -0.025] | -0.041 | 0.100 | -0.238 | 11 | (-0.075  -0.025] | -0.050 | 0.144 | -0.186 | 22 |
| (-0.025  +0.025] | -0.001 | 0.328 | -0.010 | 34 | (-0.025  +0.025] | 0.000 | 0.323 | -0.009 | 27 |
| (+0.025  +0.075] | 0.052 | 0.568 | 0.230 | 47 | (+0.025  +0.075] | 0.046 | 0.450 | 0.118 | 26 |
| (+0.075  +0.125] | 0.102 | 0.744 | 0.405 | 46 | (+0.075  +0.125] | 0.103 | 0.657 | 0.325 | 21 |
| (+0.125  +0.175] | 0.145 | 0.923 | 0.585 | 30 | (+0.125  +0.175] | 0.145 | 0.813 | 0.482 | 13 |
| (+0.175  +0.225] | 0.196 | 1.117 | 0.778 | 15 | (+0.175  +0.225] | 0.200 | 0.947 | 0.617 | 5 |
| (+0.225  +0.275] | 0.247 | 1.207 | 0.869 | 9 | (+0.225  +0.275] | 0.255 | 1.201 | 0.867 | 8 |
| (+0.275  +0.325] | 0.297 | 1.395 | 1.057 | 7 | (+0.275  +0.325] | 0.299 | 1.261 | 0.929 | 8 |
| (+0.325  +0.375] | 0.345 | 1.491 | 1.153 | 8 | (+0.325  +0.375] | 0.345 | 1.482 | 1.148 | 4 |
| | | | | | (+0.375  +0.425] | 0.398 | 1.525 | 1.196 | 4 |
| | | | | | (+0.425  +0.600] | ------ | ----- | ----- | 4 |

| (g) $0.9 < (B-V)_o \leq 1.0$ | | | | | (h) $1.0 < (B-V)_o \leq 1.1$ | | | | |
|---|---|---|---|---|---|---|---|---|---|
| $\delta_{0.6}$-interval | $<\delta_{0.6}>$ | $<\Delta M_1>$ | $<\Delta M_2>$ | N | $\delta_{0.6}$-interval | $<\delta_{0.6}>$ | $<\Delta M_1>$ | $<\Delta M_2>$ | N |
| (-0.075  -0.025] | -0.055 | 0.226 | -0.081 | 2 | (-0.325  -0.275] | ----- | ----- | ----- | 1 |
| (-0.025  +0.025] | 0.004 | 0.334 | 0.017 | 5 | (+0.025  +0.075] | 0.040 | 0.336 | 0.047 | 1 |
| (+0.025  +0.075] | 0.058 | 0.452 | 0.137 | 12 | (+0.075  +0.125] | 0.107 | 0.478 | 0.195 | 6 |
| (+0.075  +0.125] | 0.105 | 0.585 | 0.272 | 15 | (+0.125  +0.175] | 0.130 | 0.489 | 0.204 | 1 |
| (+0.125  +0.175] | 0.151 | 0.706 | 0.392 | 11 | (+0.175  +0.225] | 0.199 | 0.600 | 0.323 | 7 |
| (+0.175  +0.225] | 0.198 | 0.811 | 0.498 | 4 | (+0.225  +0.275] | 0.237 | 0.754 | 0.469 | 3 |
| (+0.275  +0.325] | 0.300 | ----- | ----- | 1 | (+0.275  +0.325] | 0.310 | 0.853 | 0.572 | 1 |

**Table 3.** Numerical values for the coefficients in equation (6) as a function of $(B-V)_o$ colour-index.

| $(B-V)_o$ | $a_3$ | $a_2$ | $a_1$ | $a_o$ |
|---|---|---|---|---|
| (0.3  0.4] | -35.7800 | +17.9170 | +1.4505 | +0.2389 |
| (0.4  0.5] | -15.4620 | +3.5129 | +4.5340 | +0.2865 |
| (0.5  0.6] | +4.9011 | -5.3226 | +5.4334 | +0.3294 |
| (0.6  0.7] | -11.0040 | -0.3570 | +5.0207 | +0.3491 |
| (0.7  0.8] | +0.0737 | -3.6154 | +4.6223 | +0.3237 |
| (0.8  0.9] | -2.4661 | +0.1822 | +3.4514 | +0.3102 |
| (0.9  1.0] | -20.8350 | +6.0860 | +2.0942 | +0.3206 |
| (1.0  1.1] | -8.3965 | +5.0002 | +1.0912 | +0.2903 |

**Table 4.** Two sets of absolute magnitudes for Hyades cluster as a function of $(B-V)_o$ colour-index. $M_V^H$: evaluated by equation (1') and $M_V^H$ (ad) adopted for normalization.

| $(B-V)_o$ | $<(B-V)_o>$ | $M_V^H$ | $M_V^H$ (ad) |
|---|---|---|---|
| (0.3  0.4] | 0.384 | 3.07 | 3.31 |
| (0.4  0.5] | 0.458 | 3.55 | 3.84 |
| (0.5  0.6] | 0.557 | 4.16 | 4.49 |
| (0.6  0.7] | 0.655 | 4.74 | 5.09 |
| (0.7  0.8] | 0.751 | 5.28 | 5.61 |
| (0.8  0.9] | 0.854 | 5.83 | 6.14 |
| (0.9  1.0] | 0.945 | 6.29 | 6.61 |
| (1.0  1.1] | 1.045 | 6.76 | 7.05 |

**Table 6.** Numerical values for the coefficients of equation (8) as a function of $(B-V)_o$ colour-index.

| $(B-V)_o$ | $b_3$ | $b_2$ | $b_1$ | $b_o$ |
|---|---|---|---|---|
| (0.3  0.4] | -32.1800 | +15.9370 | +1.7350 | -0.0177 |
| (0.4  0.5] | -15.3820 | +3.7188 | +4.4850 | +0.0022 |
| (0.5  0.6] | +3.9109 | -4.8075 | +5.3847 | +0.0134 |
| (0.6  0.7] | -11.1700 | -0.3015 | +5.0281 | +0.0153 |
| (0.7  0.8] | +0.1049 | -3.6157 | +4.6196 | -0.0144 |
| (0.8  0.9] | -22.5350 | +0.1109 | +3.4469 | -0.0203 |
| (0.9  1.0] | -24.9710 | +7.2916 | +2.0269 | +0.0051 |
| (1.0  1.1] | -7.4029 | +4.2761 | +1.2638 | -0.0047 |

**Table 7.** The mean difference between the original absolute magnitudes and the absolute magnitudes estimated by two different procedures and the corresponding standard deviations. Second and fourth columns for the new procedure and third and fifth columns for the procedure of LCL.

| | $<M_V(ori)-M_V(est)>$ | $<M_V(ori)-M_V(est)>$ | $\sigma$ | $\sigma$ |
|---|---|---|---|---|
| $(B-V)_o$ | New procedure | LCL | New procedure | LCL |
| (0.3  0.4] | +0.003 | -0.845 | ±0.087 | ±0.272 |
| (0.4  0.5] | -0.009 | -0.601 | 0.070 | 0.439 |
| (0.5  0.6] | +0.001 | -0.100 | 0.056 | 0.291 |
| (0.6  0.7] | -0.004 | +0.161 | 0.064 | 0.304 |
| (0.7  0.8] | +0.003 | +0.200 | 0.062 | 0.250 |
| (0.8  0.9] | 0.000 | +0.173 | 0.053 | 0.313 |
| (0.9  1.0] | 0.000 | +0.048 | 0.034 | 0.197 |
| (1.0  1.1] | -0.001 | -0.385 | 0.063 | 0.288 |

**Table 8.** Numerical values for the coefficients in equation (20) as a function of $(g'-r')_o$ colour-index. The $(g'-r')_o$ colour-index intervals in the first column correspond to the $(B-V)_o$ intervals in the first columns of Table 3 and Table 6.

| $(g'-r')_o$ | $c_3$ | $c_2$ | $c_1$ | $c_o$ |
|---|---|---|---|---|
| (0.09 0.19] | -12.2631 | +8.3769 | +1.2579 | -0.0177 |
| (0.19 0.30] | -5.8617 | +1.9547 | +3.2516 | +0.0022 |
| (0.30 0.40] | +1.4904 | -2.5269 | +3.9039 | +0.0134 |
| (0.40 0.51] | -4.2566 | -0.1585 | +3.6454 | +0.0153 |
| (0.51 0.61] | +0.0400 | -1.9005 | +3.3492 | -0.0144 |
| (0.61 0.72] | -0.8588 | +0.0583 | +2.4990 | -0.0203 |
| (0.72 0.82] | -9.5159 | +3.8326 | +1.4695 | +0.0051 |
| (0.82 0.93] | -2.8210 | +2.2476 | +0.9163 | -0.0047 |

**Table 9.** Colour-magnitude diagram for M13 in UBV and Sloan systems.

| $(B-V)_o$ | $M_V$ | $(g'-r')_o$ | $M_{g'}$ |
|---|---|---|---|
| 0.407 | 3.70 | 0.197 | 3.808 |
| 0.410 | 3.90 | 0.201 | 4.010 |
| 0.410 | 4.10 | 0.201 | 4.210 |
| 0.419 | 4.30 | 0.210 | 4.415 |
| 0.414 | 4.50 | 0.205 | 4.612 |
| 0.440 | 4.70 | 0.232 | 4.826 |
| 0.448 | 4.90 | 0.240 | 5.031 |
| 0.500 | 5.10 | 0.295 | 5.260 |
| 0.501 | 5.30 | 0.296 | 5.461 |
| 0.531 | 5.50 | 0.328 | 5.677 |
| 0.550 | 5.70 | 0.348 | 5.888 |
| 0.587 | 5.90 | 0.386 | 6.109 |
| 0.642 | 6.10 | 0.444 | 6.340 |
| 0.682 | 6.30 | 0.486 | 6.562 |
| 0.713 | 6.50 | 0.519 | 6.779 |
| 0.784 | 6.70 | 0.593 | 7.019 |
| 0.821 | 6.90 | 0.632 | 7.240 |
| 0.864 | 7.10 | 0.677 | 7.464 |
| 0.918 | 7.30 | 0.734 | 7.694 |
| 0.945 | 7.50 | 0.762 | 7.909 |
| 1.110 | 7.70 | 0.936 | 8.202 |

**Table 10.** Absolute magnitudes for a set of artificial stars of different metallicities with $0.2 \leq (g'-r')_o \leq 0.5$. The columns are: (1) $(g'-r')_o$ colour-index, (2) normalized absolute magnitude for a Hyades star of this colour-index, (3), (4), (5), (6), (7), (8) and (9) absolute magnitude differences $(\Delta M_{g'}^H)_{nor}$ and (10), (11), (12), (13), (14), (15) and (16) absolute magnitudes $M_{g'}$ evaluated for $\delta_{0.4}$ = 0.0, 0.1, 0.2, 0.3, 0.4, 0.5 and 0.6 respectively.

| 1 | 2 | 3 | 4 | 5 | 6 | 7 | 8 | 9 | 10 | 11 | 12 | 13 | 14 | 15 | 16 |
|---|---|---|---|---|---|---|---|---|---|---|---|---|---|---|---|
| 0.20 | 3.501 | 0.0022 | 0.341 | 0.684 | 0.995 | 1.240 | 1.384 | 1.391 | 3.483 | 3.842 | 4.185 | 4.496 | 4.742 | 4.885 | 4.892 |
| 0.21 | 3.567 | 0.0022 | 0.341 | 0.684 | 0.995 | 1.240 | 1.384 | 1.391 | 3.549 | 3.908 | 4.251 | 4.562 | 4.807 | 4.951 | 4.958 |
| 0.22 | 3.632 | 0.0022 | 0.341 | 0.684 | 0.995 | 1.240 | 1.384 | 1.391 | 3.615 | 3.973 | 4.316 | 4.628 | 4.873 | 5.016 | 5.023 |
| 0.23 | 3.697 | 0.0022 | 0.341 | 0.684 | 0.995 | 1.240 | 1.384 | 1.391 | 3.680 | 4.038 | 4.381 | 4.693 | 4.938 | 5.081 | 5.088 |
| 0.24 | 3.762 | 0.0022 | 0.341 | 0.684 | 0.995 | 1.240 | 1.384 | 1.391 | 3.744 | 4.103 | 4.446 | 4.757 | 5.003 | 5.146 | 5.153 |
| 0.25 | 3.826 | 0.0022 | 0.341 | 0.684 | 0.995 | 1.240 | 1.384 | 1.391 | 3.809 | 4.167 | 4.510 | 4.822 | 5.067 | 5.210 | 5.217 |
| 0.26 | 3.890 | 0.0022 | 0.341 | 0.684 | 0.995 | 1.240 | 1.384 | 1.391 | 3.873 | 4.231 | 4.574 | 4.886 | 5.131 | 5.274 | 5.281 |
| 0.27 | 3.954 | 0.0022 | 0.341 | 0.684 | 0.995 | 1.240 | 1.384 | 1.391 | 3.936 | 4.295 | 4.638 | 4.949 | 5.194 | 5.338 | 5.344 |
| 0.28 | 4.017 | 0.0022 | 0.341 | 0.684 | 0.995 | 1.240 | 1.384 | 1.391 | 3.999 | 4.358 | 4.701 | 5.012 | 5.257 | 5.401 | 5.408 |
| 0.29 | 4.080 | 0.0022 | 0.341 | 0.684 | 0.995 | 1.240 | 1.384 | 1.391 | 4.062 | 4.421 | 4.763 | 5.075 | 5.320 | 5.464 | 5.470 |
| 0.30 | 4.142 | 0.0022 | 0.341 | 0.684 | 0.995 | 1.240 | 1.384 | 1.391 | 4.124 | 4.483 | 4.826 | 5.137 | 5.382 | 5.526 | 5.533 |
| 0.31 | 4.204 | 0.0134 | 0.380 | 0.705 | 0.997 | 1.266 | 1.520 | 1.767 | 4.217 | 4.584 | 4.909 | 5.201 | 5.470 | 5.724 | 5.970 |
| 0.32 | 4.265 | 0.0134 | 0.380 | 0.705 | 0.997 | 1.266 | 1.520 | 1.767 | 4.279 | 4.645 | 4.970 | 5.263 | 5.531 | 5.785 | 6.032 |
| 0.33 | 4.327 | 0.0134 | 0.380 | 0.705 | 0.997 | 1.266 | 1.520 | 1.767 | 4.340 | 4.707 | 5.032 | 5.324 | 5.593 | 5.847 | 6.093 |
| 0.34 | 4.387 | 0.0134 | 0.380 | 0.705 | 0.997 | 1.266 | 1.520 | 1.767 | 4.401 | 4.767 | 5.092 | 5.385 | 5.653 | 5.907 | 6.154 |
| 0.35 | 4.448 | 0.0134 | 0.380 | 0.705 | 0.997 | 1.266 | 1.520 | 1.767 | 4.461 | 4.828 | 5.153 | 5.445 | 5.714 | 5.968 | 6.214 |
| 0.36 | 4.508 | 0.0134 | 0.380 | 0.705 | 0.997 | 1.266 | 1.520 | 1.767 | 4.521 | 4.888 | 5.213 | 5.505 | 5.774 | 6.028 | 6.274 |
| 0.37 | 4.567 | 0.0134 | 0.380 | 0.705 | 0.997 | 1.266 | 1.520 | 1.767 | 4.581 | 4.947 | 5.272 | 5.565 | 5.833 | 6.087 | 6.334 |
| 0.38 | 4.627 | 0.0134 | 0.380 | 0.705 | 0.997 | 1.266 | 1.520 | 1.767 | 4.640 | 5.007 | 5.332 | 5.624 | 5.893 | 6.147 | 6.393 |
| 0.39 | 4.686 | 0.0134 | 0.380 | 0.705 | 0.997 | 1.266 | 1.520 | 1.767 | 4.699 | 5.066 | 5.391 | 5.683 | 5.952 | 6.205 | 6.452 |
| 0.40 | 4.744 | 0.0134 | 0.380 | 0.705 | 0.997 | 1.266 | 1.520 | 1.767 | 4.757 | 5.124 | 5.449 | 5.741 | 6.010 | 6.264 | 6.511 |
| 0.41 | 4.802 | 0.0153 | 0.374 | 0.704 | 0.980 | 1.176 | 1.266 | 1.224 | 4.817 | 5.176 | 5.506 | 5.782 | 5.978 | 6.068 | 6.026 |
| 0.42 | 4.860 | 0.0153 | 0.374 | 0.704 | 0.980 | 1.176 | 1.266 | 1.224 | 4.875 | 5.234 | 5.564 | 5.840 | 6.035 | 6.126 | 6.084 |
| 0.43 | 4.917 | 0.0153 | 0.374 | 0.704 | 0.980 | 1.176 | 1.266 | 1.224 | 4.932 | 5.291 | 5.621 | 5.897 | 6.093 | 6.183 | 6.141 |
| 0.44 | 4.974 | 0.0153 | 0.374 | 0.704 | 0.980 | 1.176 | 1.266 | 1.224 | 4.989 | 5.348 | 5.678 | 5.954 | 6.150 | 6.240 | 6.198 |
| 0.45 | 5.031 | 0.0153 | 0.374 | 0.704 | 0.980 | 1.176 | 1.266 | 1.224 | 5.046 | 5.405 | 5.735 | 6.010 | 6.206 | 6.297 | 6.255 |
| 0.46 | 5.087 | 0.0153 | 0.374 | 0.704 | 0.980 | 1.176 | 1.266 | 1.224 | 5.102 | 5.461 | 5.791 | 6.066 | 6.262 | 6.353 | 6.311 |
| 0.47 | 5.142 | 0.0153 | 0.374 | 0.704 | 0.980 | 1.176 | 1.266 | 1.224 | 5.158 | 5.516 | 5.846 | 6.122 | 6.318 | 6.409 | 6.366 |
| 0.48 | 5.198 | 0.0153 | 0.374 | 0.704 | 0.980 | 1.176 | 1.266 | 1.224 | 5.213 | 5.572 | 5.902 | 6.178 | 6.374 | 6.464 | 6.422 |
| 0.49 | 5.253 | 0.0153 | 0.374 | 0.704 | 0.980 | 1.176 | 1.266 | 1.224 | 5.268 | 5.627 | 5.957 | 6.233 | 6.429 | 6.519 | 6.477 |
| 0.50 | 5.307 | 0.0153 | 0.374 | 0.704 | 0.980 | 1.176 | 1.266 | 1.224 | 5.323 | 5.681 | 6.011 | 6.287 | 6.483 | 6.574 | 6.531 |

**Table 11.** Comparison of the absolute magnitudes estimated by the new procedure and by means of colour-magnitude diagram for cluster M13 for artificial stars in question. The columns give: (1) colour index $(g'-r')_o$ (2) absolute magnitude $M_{g'}(M13)$ evaluated by equation (22), (3), (4), (5), (6), (7), (8) and (9) difference between the absolute magnitude $M_{g'}(M13)$ and the absolute magnitudes estimated for $\delta_{0.4} = 0.0, 0.1, 0.2, 0.3, 0.4, 0.5$, and 0.6 respectively (columns 10-16, Table 10). Averages of these differences ($<\Delta M>$) and the corresponding standard deviations ($\sigma$) are given in the last two lines.

| 1 | 2 | 3 | 4 | 5 | 6 | 7 | 8 | 9 |
|---|---|---|---|---|---|---|---|---|
| 0.20 | 4.260 | 0.776 | 0.418 | 0.075 | -0.237 | -0.482 | -0.625 | -0.632 |
| 0.21 | 4.386 | 0.837 | 0.478 | 0.136 | -0.176 | -0.421 | -0.565 | -0.571 |
| 0.22 | 4.509 | 0.895 | 0.536 | 0.193 | -0.118 | -0.363 | -0.507 | -0.514 |
| 0.23 | 4.629 | 0.949 | 0.590 | 0.248 | -0.064 | -0.309 | -0.453 | -0.459 |
| 0.24 | 4.745 | 1.000 | 0.642 | 0.299 | -0.013 | -0.258 | -0.401 | -0.408 |
| 0.25 | 4.857 | 1.049 | 0.690 | 0.347 | 0.036 | -0.209 | -0.353 | -0.360 |
| 0.26 | 4.967 | 1.094 | 0.735 | 0.393 | 0.081 | -0.164 | -0.308 | -0.314 |
| 0.27 | 5.073 | 1.137 | 0.778 | 0.435 | 0.124 | -0.122 | -0.265 | -0.272 |
| 0.28 | 5.175 | 1.176 | 0.817 | 0.475 | 0.163 | -0.082 | -0.225 | -0.232 |
| 0.29 | 5.275 | 1.213 | 0.854 | 0.512 | 0.200 | -0.045 | -0.189 | -0.195 |
| 0.30 | 5.371 | 1.247 | 0.889 | 0.546 | 0.234 | -0.011 | -0.154 | -0.161 |
| 0.31 | 5.465 | 1.248 | 0.881 | 0.556 | 0.264 | -0.005 | -0.259 | -0.505 |
| 0.32 | 5.556 | 1.277 | 0.910 | 0.585 | 0.293 | 0.024 | -0.230 | -0.476 |
| 0.33 | 5.643 | 1.303 | 0.937 | 0.612 | 0.319 | 0.051 | -0.203 | -0.450 |
| 0.34 | 5.728 | 1.328 | 0.961 | 0.636 | 0.344 | 0.075 | -0.179 | -0.426 |
| 0.35 | 5.811 | 1.349 | 0.983 | 0.658 | 0.365 | 0.097 | -0.157 | -0.404 |
| 0.36 | 5.890 | 1.369 | 1.002 | 0.677 | 0.385 | 0.116 | -0.138 | -0.384 |
| 0.37 | 5.967 | 1.386 | 1.020 | 0.695 | 0.402 | 0.134 | -0.120 | -0.367 |
| 0.38 | 6.042 | 1.402 | 1.035 | 0.710 | 0.418 | 0.149 | -0.105 | -0.352 |
| 0.39 | 6.114 | 1.415 | 1.048 | 0.723 | 0.431 | 0.162 | -0.092 | -0.339 |
| 0.40 | 6.183 | 1.426 | 1.059 | 0.734 | 0.442 | 0.173 | -0.081 | -0.327 |
| 0.41 | 6.251 | 1.433 | 1.075 | 0.745 | 0.469 | 0.273 | 0.182 | 0.225 |
| 0.42 | 6.316 | 1.441 | 1.082 | 0.752 | 0.476 | 0.280 | 0.190 | 0.232 |
| 0.43 | 6.379 | 1.446 | 1.088 | 0.758 | 0.482 | 0.286 | 0.195 | 0.238 |
| 0.44 | 6.439 | 1.450 | 1.091 | 0.761 | 0.486 | 0.290 | 0.199 | 0.241 |
| 0.45 | 6.498 | 1.452 | 1.094 | 0.764 | 0.488 | 0.292 | 0.201 | 0.244 |
| 0.46 | 6.555 | 1.453 | 1.094 | 0.764 | 0.489 | 0.293 | 0.202 | 0.244 |
| 0.47 | 6.610 | 1.452 | 1.093 | 0.763 | 0.488 | 0.292 | 0.201 | 0.243 |
| 0.48 | 6.663 | 1.450 | 1.091 | 0.761 | 0.485 | 0.289 | 0.199 | 0.241 |
| 0.49 | 6.714 | 1.446 | 1.087 | 0.757 | 0.482 | 0.286 | 0.195 | 0.237 |
| 0.50 | 6.764 | 1.441 | 1.082 | 0.752 | 0.477 | 0.281 | 0.190 | 0.232 |
| $<\Delta M>$ | | 1.269 | 0.908 | 0.575 | 0.281 | 0.044 | -0.118 | -0.186 |
| $\sigma$ | | 0.205 | 0.204 | 0.210 | 0.224 | 0.237 | 0.256 | 0.313 |

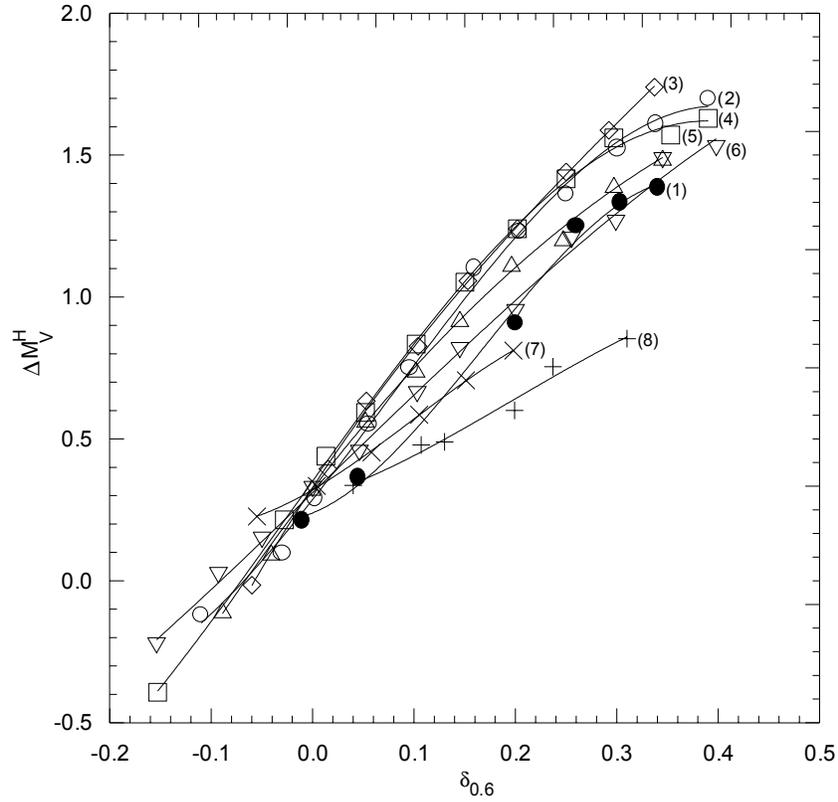

**Fig.1.** $\Delta M_V^H$ versus $\delta_{0.6}$ for eight $(B-V)_o$ colour-index intervals. The symbols give (•): $0.3 < (B-V)_o \leq 0.4$, (O): $0.4 < (B-V)_o \leq 0.5$, (◊): $0.5 < (B-V)_o \leq 0.6$, (□): $0.6 < (B-V)_o \leq 0.7$, (Δ): $0.7 < (B-V)_o \leq 0.8$, (∇): $0.8 < (B-V)_o \leq 0.9$, (X): $0.9 < (B-V)_o \leq 1.0$, and (+): $1.0 < (B-V)_o \leq 1.1$.

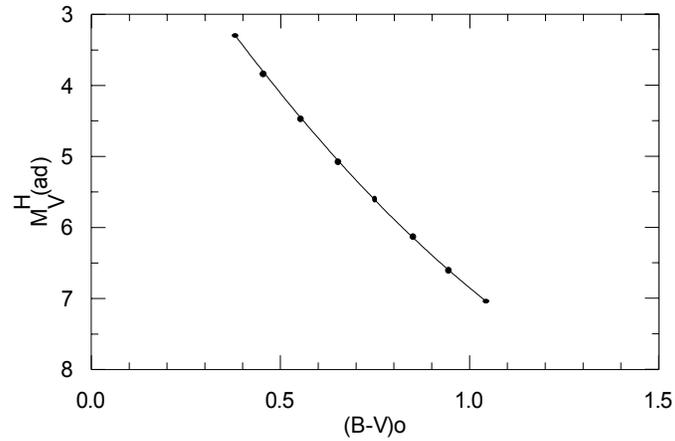

**Fig.2.** Adopted absolute magnitudes for the Hyades main-sequence versus $(B-V)_o$ colour-index, presented.

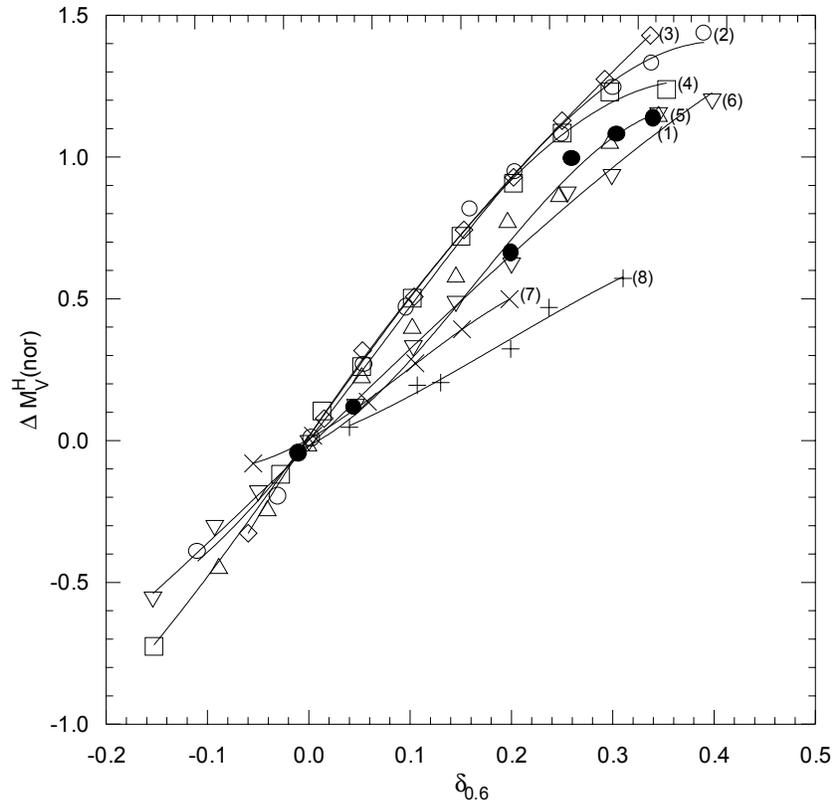

**Fig. 3.** $\Delta M_V^H(nor)$ versus $\delta_{0.6}$ for eight $(B-V)_o$ colour-index intervals (symbols as in Fig.1).

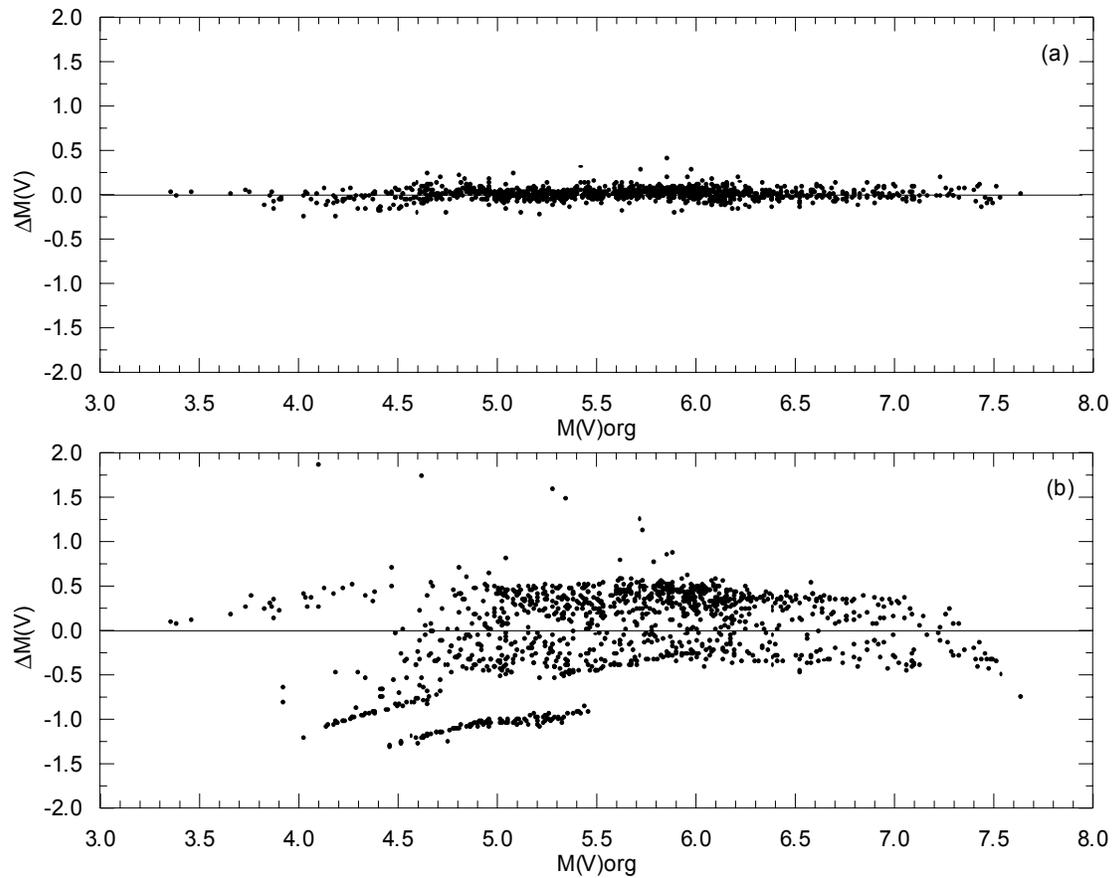

**Fig.4.** Deviation of the evaluated absolute magnitudes relative to the original absolute magnitudes. (a) for the new procedure, and (b) for LCL.

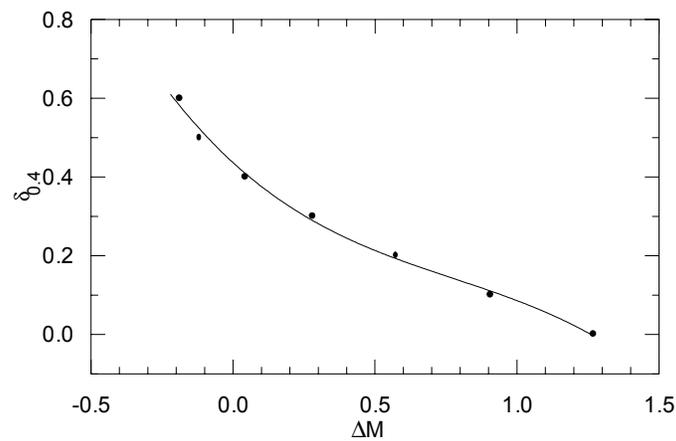

**Fig.5.** $\delta_{0.4}$ versus mean absolute magnitude difference $\langle\Delta M\rangle$.

**Appendix**

We used the update data for a Hyades sequence evaluation. 46 stars ($\sigma_\pi/\pi$ <0.12) within the tidal radius, r(t)≤ 10 pc, of the cluster are adopted as the cluster members and their absolute magnitudes are evaluated by their parallaxes and the V apparent magnitudes. Then a second degree polynomial has been fitted between M(V) and (B-V), i.e.:

$$M(V) = -1.48739(B-V)^2 + 7.70982(B-V) + 0.331195 \qquad (A.1)$$

(Fig.A.1.). The data in Table A.1 are taken from two different sources, due to the lack of (U-B) in Hipparcos catalogue. The parallax data and the colour indices $(B-V)_H$ are provided from Perryman (private communication, see also Perryman et al. (1998)), whereas V, (B-V), and (U-B) are taken from SIMBAD database. Fig.A.2. shows that there is no any conspicuous difference between two set of (B-V) data.

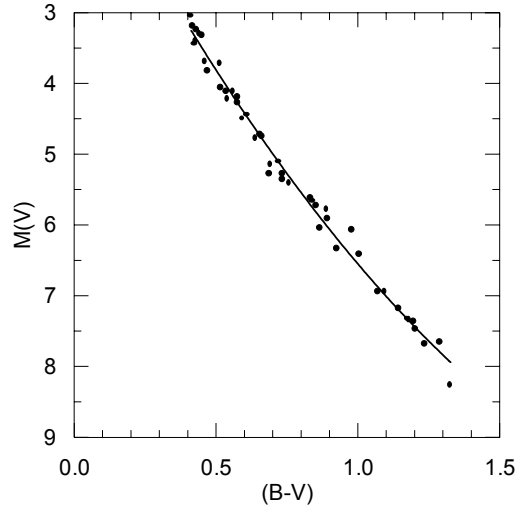
**Fig.A.1.** Update sequence for Hyades cluster.

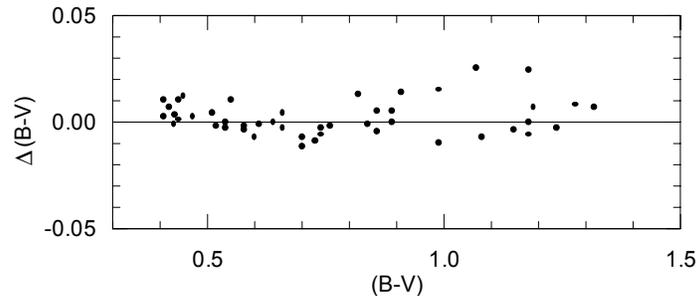
**Fig.A.2.** Comparison of the B-V colour indices for 46 Hyades member taken from two different sources: Perryman (private communication) and SIMBAD database.

**Table A.1.** Data for Hyades cluster. The columns give ID number, Hipparcos number, parallax, standard error for the parallax, absolute magnitude, (B-V)$_H$ colour index, V apparent magnitude, B-V and U-B indices.

| ID | Hip No | $\pi$ | $\sigma_\pi/\pi$ | M(V) | (B-V)$_H$ | V | (B-V) | (U-B) |
|---|---|---|---|---|---|---|---|---|
| 1 | 20357 | 19.46 | 1.02 | 3.05 | 0.412 | 6.61 | 0.41 | 0.00 |
| 2 | 21152 | 23.13 | 0.92 | 3.19 | 0.420 | 6.40 | 0.41 | 0.00 |
| 3 | 20349 | 19.55 | 0.89 | 3.25 | 0.434 | 6.80 | 0.43 | -0.03 |
| 4 | 20350 | 19.83 | 0.89 | 3.29 | 0.441 | 6.81 | 0.44 | 0.00 |
| 5 | 20567 | 18.74 | 1.17 | 3.32 | 0.450 | 6.97 | 0.44 | 0.00 |
| 6 | 21267 | 22.80 | 0.98 | 3.41 | 0.429 | 6.62 | 0.43 | -0.01 |
| 7 | 19504 | 23.22 | 0.92 | 3.44 | 0.427 | 6.62 | 0.42 | -0.01 |
| 8 | 20491 | 20.04 | 0.89 | 3.69 | 0.462 | 7.18 | 0.45 | 0.03 |
| 9 | 19796 | 21.08 | 0.97 | 3.73 | 0.514 | 7.12 | 0.51 | 0.05 |
| 10 | 21066 | 22.96 | 0.99 | 3.83 | 0.472 | 7.04 | 0.47 | 0.00 |
| 11 | 20557 | 24.47 | 1.06 | 4.07 | 0.518 | 7.14 | 0.52 | 0.04 |
| 12 | 20815 | 21.83 | 1.01 | 4.11 | 0.537 | 7.42 | 0.54 | 0.06 |
| 13 | 20826 | 21.18 | 1.04 | 4.12 | 0.560 | 7.51 | 0.55 | 0.05 |
| 14 | 22422 | 19.68 | 0.96 | 4.19 | 0.578 | 7.74 | 0.58 | 0.11 |
| 15 | 21112 | 19.46 | 1.02 | 4.23 | 0.540 | 7.77 | 0.54 | 0.06 |
| 16 | 21637 | 22.60 | 0.91 | 4.28 | 0.576 | 7.53 | 0.58 | 0.10 |
| 17 | 20899 | 21.09 | 1.08 | 4.45 | 0.609 | 7.85 | 0.61 | 0.13 |
| 18 | 19148 | 21.41 | 1.47 | 4.50 | 0.593 | 7.85 | 0.60 | 0.10 |
| 19 | 19793 | 21.69 | 1.14 | 4.73 | 0.657 | 8.09 | 0.66 | 0.20 |
| 20 | 20741 | 21.42 | 1.54 | 4.75 | 0.664 | 8.12 | 0.66 | 0.20 |
| 21 | 19786 | 22.19 | 1.45 | 4.78 | 0.640 | 8.06 | 0.64 | 0.17 |
| 22 | 20146 | 21.24 | 1.32 | 5.11 | 0.721 | 8.45 | 0.73 | 0.31 |
| 23 | 19781 | 21.91 | 1.27 | 5.15 | 0.693 | 8.46 | 0.70 | 0.24 |
| 24 | 21099 | 21.81 | 1.25 | 5.28 | 0.734 | 8.60 | 0.74 | 0.35 |
| 25 | 21788 | 19.48 | 1.26 | 5.29 | 0.688 | 8.83 | 0.70 | 0.24 |
| 26 | 23069 | 19.66 | 1.62 | 5.36 | 0.737 | 8.89 | 0.74 | 0.35 |
| 27 | 20480 | 20.63 | 1.34 | 5.41 | 0.758 | 8.85 | 0.76 | 0.35 |
| 28 | 22380 | 21.38 | 1.46 | 5.63 | 0.833 | 8.97 | 0.82 | 0.51 |
| 29 | 20850 | 21.29 | 1.91 | 5.66 | 0.839 | 9.04 | 0.84 | 0.53 |
| 30 | 20492 | 21.23 | 1.80 | 5.74 | 0.855 | 9.12 | 0.86 | 0.54 |
| 31 | 19098 | 19.81 | 1.39 | 5.79 | 0.890 | 9.29 | 0.89 | 0.58 |
| 32 | 18327 | 24.16 | 1.40 | 5.91 | 0.895 | 8.99 | 0.89 | 0.64 |
| 33 | 20978 | 24.71 | 1.27 | 6.04 | 0.865 | 9.11 | 0.86 | 0.56 |
| 34 | 20082 | 20.01 | 1.91 | 6.08 | 0.980 | 9.60 | 0.99 | 0.80 |
| 35 | 20187 | 20.13 | 2.02 | 6.33 | 0.924 | 9.81 | 0.91 | |
| 36 | 19263 | 19.70 | 1.68 | 6.41 | 1.005 | 9.94 | 0.99 | 0.89 |
| 37 | 18946 | 23.07 | 2.12 | 6.94 | 1.095 | 10.13 | 1.07 | 1.03 |
| 38 | 21723 | 23.95 | 1.63 | 6.94 | 1.073 | 10.04 | 1.08 | 1.00 |
| 39 | 20762 | 21.83 | 2.29 | 7.18 | 1.146 | 10.44 | 1.15 | 1.04 |
| 40 | 22271 | 22.07 | 2.03 | 7.33 | 1.174 | 10.60 | 1.18 | 1.10 |
| 41 | 19207 | 23.57 | 2.26 | 7.35 | 1.180 | 10.48 | 1.18 | 1.12 |
| 42 | 21261 | 21.06 | 2.21 | 7.36 | 1.197 | 10.74 | 1.19 | 1.17 |
| 43 | 19808 | 22.67 | 2.30 | 7.47 | 1.204 | 10.75 | 1.18 | 1.18 |
| 44 | 20527 | 22.57 | 2.78 | 7.66 | 1.288 | 10.90 | 1.28 | 1.26 |
| 45 | 21256 | 24.98 | 1.95 | 7.68 | 1.237 | 10.67 | 1.24 | 1.13 |
| 46 | 19316 | 24.90 | 2.59 | 8.26 | 1.327 | 11.27 | 1.32 | 1.26 |